\documentstyle[12pt,graphicx]{article}
\title{ Relic Gravitational Waves in the  Accelerating Universe
      }
\author{Yang  Zhang, Yefei Yuan,  Wen  Zhao, and Ying-Tian Chen \\
        Astrophysics Center \\
        University of Science and Technology of China \\
        Hefei, Anhui, China }
 \date{}

\begin{document}
\maketitle
\baselineskip=19truept
\def\vek{\vec{k}}

\newcommand{\be}{\begin{equation}}
\newcommand{\ee}{\end{equation}}
\newcommand{\ba}{\begin{eqnarray}}
\newcommand{\ea}{\end{eqnarray}}

\sf

\begin{center}
\Large  Abstract
\end{center}
\begin{quote}
 {\large
Observations have recently indicated that the Universe at the
present stage is in an accelerating expansion, a process that has
great implications.
 We  evaluate the spectrum of relic
gravitational waves in the current accelerating Universe
and find  that there are new features appearing in  the resulting
spectrum as  compared  to the decelerating models. In the low
frequency range  the peak of
spectrum is now located at a frequency $\nu_E \simeq
(\frac{\Omega_m}{\Omega_{\Lambda}})^{1/3} \nu_H$,
where $\nu_H$ is the Hubble frequency,
and there
appears a new segment of spectrum between  $\nu_E$ and $\nu_H$. In
all other intervals of frequencies $\geq  \nu_H$,
the spectral amplitude  acquires an extra
factor $\frac{\Omega_m}{\Omega_{\Lambda}}$, due to the current
acceleration, otherwise the shape of spectrum is similar to
that in the decelerating models.
The recent WMAP result of CMB anisotropies is used
to normalize the amplitude for gravitational waves.
The slope of the power
spectrum depends sensitively on the scale factor $a(\tau) \propto
|\tau|^{1+\beta}$ during the inflationary stage
 with $\beta=-2$ for the exact de Sitter space.
 With the increasing  of $\beta$,
the resulting spectrum is tilted to be flatter with  more power  on high
frequencies, and the sensitivity of the second science run of the LIGO detectors
puts a restriction on the parameter $\beta \leq -1.8$.
We also give a numerical solution which confirms these features.
 }
\end{quote}



e-mail: yzh@ustc.edu.cn

\newpage
\baselineskip=19truept

\begin{center}
{\em\Large 1. Introduction.}
\end{center}

The inflationary expansion of the early Universe can create a
stochastic background of relic gravitational waves, which is
important in cosmology and has been extensively studied in the
past \cite{starobinsky} \cite{rubakov} \cite{fabbri} \cite{abbott}
\cite{harari}. The spectrum of relic gravitational waves, as is to
be observed today, depends not only on the details of the early
stage of inflationary expansion, but also on the expansion
behavior of the subsequent stages, including the current one.  The
calculations of spectrum  so far \cite{allen} \cite{sahni}
\cite{grishchuk} \cite{riazuelo} \cite{tashiro}
\cite{henriques} have been done for the case that the current
stage is in a decelerating expansion. The resulting spectrum has
been put among the candidate list of sources for  the
gravitational wave detectors either in operation
\cite{barish}\cite{willke}, or  under construction \cite{bender}
\cite{larson}. The astronomical observations on SN Ia \cite{riess}
\cite{perlmutter} indicate that the Universe  is currently under
accelerating expansion, which may be driven by the cosmic dark
energy ($\Omega_{\Lambda} \sim 0.7$) plus the dark matter
($\Omega_{m} \sim 0.3$) \cite{bahcall}.
  This is further supported by the  recent
WMAP results on CMB anisotropies \cite{spergel} with $\Omega =1$.
By the
wave equation, the evolution of relic gravitational waves
depends on the expanding spacetime background, and the wave
amplitudes depend on whether the wave lengths  are inside
or outside the Hubble radius.
The Universe under accelerating expansion has a Hubble radius
as a function of time that differs from that in the
conventional models of decelerating expansion. Therefore,
one can expect that, if an earlier matter-dominated stage
of decelerating expansion is followed by the current
accelerating expansion, the outcome for  the spectrum of
relic gravitational waves will be altered.

In this paper we study the impact of the current accelerating
expansion on the relic gravitational waves.
We will first sketch,
as a setup, the well-known formulations of the gravitational waves
in an expanding spacetime \cite{grish}, and give the explicit
scale factor $a(\tau)$ for a sequence of successive expanding epochs,
including the current  epoch of accelerating expansion.
We then evaluate the power spectrum of relic gravitational waves and find that
the current accelerating expansion does change the spectrum,
including its shape and amplitude.
Throughout the paper we shall work with units with $c=\hbar=1$,
otherwise it will be pointed out. We also use  notations similar
to that of Grishchuk \cite{grishchuk} for convenience for
comparison.

\begin{center}
{\em\Large 2. The Gravitational Wave Equation.}
\end{center}

Incorporating the perturbations to the spatially flat
Robertson-Walker space-time, the metric is
\be
ds^2=a^2(\tau) [  d\tau^2-(\delta_{ij}+h_{ij})dx^idx^j ],
\ee
where where $\tau $ is the conformal  time, the
perturbations of spacetime $h_{ij}$ is a $3\times 3$
symmetric matrix   containing  generally the scalar,
vector, and tensor parts. The gravitational wave field is
the tensorial portion of $h_{ij}$, which is
transverse-traceless
\be
\partial_i h^{ij}=0,\,\,\,\,
\delta^{ij}h_{ij}=0.
\ee
We are interested only in the creation of the  relic gravitational waves
by the expanding spacetime background,
 the perturbed matter source is therefore not taken into account.
Moreover, as the relic gravitational waves are very weak,
in the sense that $h_{ij}\ll 1$,
so one need just  study the linearized field equation:
\be
\partial_{\mu}(\sqrt{-g}\partial^{\mu}h_{ij}({\bf{x}} ,\tau))=0 .
\ee
In quantum theory of gravitational waves, the field
$h_{ij}$ is a field operator, which is written as a sum of
 the plane wave Fourier modes
\be
 h_{ij}({\bf{x}},\tau) \frac{  \sqrt{16\pi}l_{Pl}   }{(2\pi)^{\frac{3}{2}}} \sum_{\lambda=1}^2
\int_{-\infty}^{\infty} d^3{\bf k} \epsilon^{(\lambda)}_{ij}({\bf
k}) \frac{1}{\sqrt{2k}} [ a_k^{(\lambda)}  h^{(\lambda)}_k(\tau)
e^{i\bf k\cdot x} + a_k^{\dagger(\lambda)} h^{(\lambda)
*}_k(\tau)e^{-i\bf k\cdot x} ],
\ee
where $l_{Pl} = \sqrt{G}$ is the Planck's length,
the two  polarizations $\epsilon^{(\lambda)}_{ij}(\bf k)$, $\lambda =1, 2$,
are symmetric and transverse-traceless
$   k^i\epsilon^{(\lambda)}_{ij}({\bf k})=0$,
$ \delta^{ij} \epsilon^{(\lambda)}_{ij}({\bf k})=0  $,
and satisfy the conditions
$  \epsilon ^{(\lambda)ij}({\bf k}) \epsilon^{(\lambda')}_{ij}({\bf k})
=  2\delta_{\lambda\lambda'} $,
and
$\epsilon^{(\lambda)}_{ij}({\bf -k}) = \epsilon^{(\lambda)}_{ij}({\bf k})$,
the creation and annihilation operators satisfy
$ [ a_{\bf k}^{(\lambda)}  , a_{\bf k'}^{\dagger(\lambda')} ]
   = \delta_{\lambda\lambda'}\delta^3({\bf k}-{\bf k'}) $,
and  the initial vacuum state is  defined as \be a_{\bf
k}^{(\lambda)}  \mid 0> = 0 \ee for each $\bf k$ and
$\lambda$. As a matter of fact, this definition depends on
the choice of the mode function $  h^{(\lambda)}_k(\tau) $;
different $  h^{(\lambda)}_k(\tau) $ define  different
vacuum states, a point that  will be further explained
later. In the vacuum state the energy density of
gravitational waves $ t_{00} = \frac{1}{32\pi G} h_{ij,\,
0} h^{ij}_{\,\, ,\, 0} $ gives the zero-point energy
\[
<0| \int t_{00}d^3{\bf x}  |0>
= \frac{1}{2}\hbar
\int^{\infty}_{ -\infty}d^3{\bf k}
\sum_{\lambda = 1}^{2}
\omega_k
<0| a^{(\lambda)}_{\bf k}   a^{(\lambda)\dagger}_{\bf k}  |0>.
\]

For  a fixed wave number $\bf k$ and a fixed polarization
state $\lambda$, the Eq.(3) reduces to the second-order
ordinary differential  equation \cite{grish}
\be
 h_k^{(\lambda)''}(\tau)
+2\frac{a'}{a}h_k^{(\lambda)'} (\tau)+k^2h^{(\lambda)} _k
=0,
\ee
 where the prime denotes $d/d\tau$. Since the equation of $h_{\bf k}^{(\lambda)} (\tau) $
 for each polarization is the same, we denote $h_{\bf k}^{(\lambda)}
(\tau) $ by $h_{\bf k}(\tau) $. One rescales the  field
$h_{\bf k} (\tau) $ as
\be h_{\bf k} (\tau) \frac{\mu_{\bf k}(\tau)}{a(\tau)},
\ee
and the equation for
$\mu_{\bf k}$ is
\be \mu_k'' + (k^2-\frac{a''}{a} )\mu_k
=0.
\ee
This equation can be regarded as the equation for
one-dimensional oscillator in a given effective potential
barrier $a''/a$. For a given spacetime background with a
generic power-law form of the scale factor
\be
a(\tau) \propto \tau^{\alpha},
\ee
where $\alpha$ is a real number,
positive or negative, the general solution is a linear
combination of the Hankel's functions:
\be
\mu_{ k}(\tau) A_k \sqrt{k\tau} H^{(1)}_{ (\alpha-\frac{1}{2})} (k\tau)
+B_k \sqrt{k\tau}H^{(2)}_{ (\alpha-\frac{1}{2})}(k\tau).
\ee
Given a model of the expansion of Universe, consisting
of a sequence of successive $a(\tau)$ as in Eq.(9)
with  different $\alpha$, one can construct an exact solution
$\mu_k(\tau)$ by matching its values and its derivatives at
the joining points.
One may may also  numerically solve  Eq.(6)
and give the corresponding  power spectrum as we will present later in the paper.
The vacuum state $|0>$, determined by
the mode function $h_{ k}(\tau) $, is therefore fixed by a
choice of coefficients $A_k$ and $B_k$. In the limit
$k\rightarrow \infty$, or $\tau \rightarrow  \pm \infty$,
\[
\sqrt{k\tau}  H^{(1)}_{(\alpha-\frac{1}{2})}(k\tau)
\rightarrow \sqrt{\frac{2}{\pi}}i^{-\alpha} e^{-ik\tau},
\]
\[
\sqrt{k\tau}  H^{(2)}_{(\alpha-\frac{1}{2})}(k\tau)
\rightarrow \sqrt{\frac{2}{\pi}}i^{\alpha} e^{ik\tau},
\]
approaching the positive and negative frequency modes,
respectively. For instance, if we choose  $B_k=0$, then in the
limit $k\rightarrow \infty$, or $\tau \rightarrow  \pm \infty$,
\[
h_{\bf k}  (\tau) \propto e^{-ik\tau},
\]
giving the positive frequency mode.
This choice yields the so-called adiabatic vacuum \cite{parker} through (4) and (5).
The amplitude of the power spectrum of the gravitational waves depends only on the
initial value of $|h_{\bf k}  (\tau)|$ at the horizon-crossing.

The following  two limiting cases
are useful for an approximate evaluation of the spectrum.
Outside the barrier $k^2 \gg \frac{a''}{a}$ (equivalent to  $k\gg \frac{a'}{a}$)
the gravitational wave field reduces to
\be
h_{ k}(\tau) \rightarrow  A_k\frac{e^{ik\tau}}{a(\tau)}+ B_k \frac{e^{-ik\tau}}{a(\tau)},
\ee
having a decreasing amplitude
\be
h_k(\tau) \propto 1/a(\tau).
\ee
Inside the barrier  $k^2 \ll \frac{a''}{a}$ (equivalent to  $k\ll \frac{a'}{a}$)
the gravitational wave field
\be
h_{ k}(\tau) \rightarrow  C_k + D_k  \int^{\tau}\frac{d\tau'}{a^2(\tau')}.
\ee
The second term $D\int^{\tau}\frac{d\tau'}{a^2(\tau')}$  is small
for the models that we shall study in the following,
so the long wave-length limit of $h_k$ is simply a constant:
\be
h_k(\tau) = C_k.
\ee
Thus, as a function of $\tau$, $h_k(\tau)$ has simple approximate behaviors
in the two limiting cases, and we will use these to
estimate the spectrum  at present stage.

\begin{center}
{\bf\Large  3. Epochs of Expanding Universe}
\end{center}

The history of overall expansion of the Universe can be  modelled
as the following sequence of successive epochs of power-law expansion.

The initial stage (inflationary)
\be
a(\tau ) = l_0 \mid {\tau} \mid ^{1+\beta},  \,\,\,\,\, -\infty < \tau \leq  \tau_1,
\ee
where $1+\beta<0$, and $\tau_1<0$. The special case of $\beta=-2$
is the de Sitter expansion.

The  z-stage
\be
a(\tau) = a_z(\tau-  \tau_p)^{1+\beta_s}, \,\,\,\,\,  \tau_1 \leq  \tau \leq  \tau_s,
\ee
where  $\beta_s +1 >0$.
Towards the end of inflation, during the reheating,
the equation of state of energy in the Universe can be quite complicated
and is rather  model-dependent  \cite{ford}.
So this z-stage is introduced to allow a general reheating epoch,
as has been advocated by Grishchuk \cite{grishchuk}.

The radiation-dominated stage
\be
a(\tau) = a_e(\tau  -\tau_e), \,\,\,\,\, \tau_s \leq  \tau \leq  \tau_2.
\ee

The matter-dominated stage
\be
a(\tau) =  a_m(\tau  -\tau_m)^2 , \,\,\,\,\,  \tau_2  \leq  \tau \leq  \tau_E,
\ee
where $\tau_E$ is the time when the dark energy density $\rho_{\Lambda}$
is equal to the matter energy density $\rho_m$.
Before the discovery of the  accelerating expansion of the Universe,
the current expansion was usually taken to be in this matter-dominated
stage, which is a decelerating one.
Now, following the matter-dominated stage, we add an epoch of accelerating stage .
The value of the redshift $z_E$ at the time $\tau_E$ is given by
$ (1+z_E) = a(\tau_H)/a(\tau_E) $,
where $\tau_H$ is the present time.
Since $\rho_{\Lambda}$ is constant
and $\rho_m(\tau) \propto a^{-3}(\tau)$,
 one has
\[
1=\frac{\rho_{\Lambda}}{\rho_m(\tau_E)}=\frac{\rho_{\Lambda}}{\rho_m(\tau_H)(1+z_E)^3}.
\]
If we take the current values $\Omega_{\Lambda} \sim  0.7$ and $\Omega_m \sim  0.3$,
then it follows that
\[
1+z_E = (\frac{\Omega_{\Lambda}}{\Omega_m})^{1/3} \sim 1.33.
\]

The accelerating stage (up to the present)
\be
a(\tau) =  l_H \mid \tau-  \tau_a\mid ^{-1},
\,\,\,\,\,  \tau_E \leq  \tau  \leq   \tau_H .
\ee
This stage describes the accelerating expansion of the Universe,
which is the new feature in our model and will induce some modifications
to the spectrum of the relic gravitational waves.
It should be mentioned that the actual scale factor function $a(\tau)$ differs from Eq.(19),
since the matter component  exists in the current Universe.
However, the dark energy is dominant,
so (19) is an approximation  to the current expansion behavior.

Given $a(\tau)$ for the various epochs,
the derivative $a'=da/d\tau$ and the ratio $a'/a$ follow immediately.
Except for the $\beta_s$ that is imposed upon as the model parameter,
there are ten constants in the above expressions of $a(\tau)$.
By the continuity conditions  of $a(\tau)$ and  $a(\tau)'$ at the  four given
joining points $\tau_1$, $\tau_s$, $\tau_2$, and $\tau_E$,
one can fix only eight constants.
The remaining two  constants can be fixed by
the overall normalization of $a$ and by the observed Hubble constant as the expansion rate.
Specifically,  we put $\mid \tau_H  -   \tau_a \mid = 1$
as the normalization of $a$, which fixes the constant $\tau_a$,
 and the constant $l_H  $ is fixed by the following calculation
\be
\frac{1}{H} \equiv  \left(\frac{a^2}{a'}   \right)_{\tau_H}  = l_H  ,
\ee
so $l_H$ is just the  Hubble radius  at present.
Then everything is fixed up.
In the expanding Robertson-Walker spacetime
the physical wave length is related to the comoving wave number by
\be
\lambda
\equiv \frac{2\pi a(\tau)}{k},
\ee
and the wave number $k_H$ corresponding to the present Hubble radius is
\be
 k_H = \frac{2\pi a(\tau_H )}{l_H}  =2\pi .
\ee
There is another wave number,
\[
k_E \equiv \frac{2\pi a(\tau_E )}{1/H}=\frac{k_H}{1+z_E},
\]
 whose corresponding wave length at the time $\tau_E$ is the
the Hubble radius $1/H$.

By matching $a$ and $a'/a$ at the joint points,
we have  derived, for example,
\be
l_0  =  l_Hb
\zeta_E^{-(2+\beta)} \zeta_2^{\frac{\beta-1}{2}} \zeta_s^{\beta}
\zeta_1^{\frac{\beta-\beta_s}{1+\beta_s}},
\ee
where $b \equiv |1+\beta|^{-(1+\beta)}$,
which is defined differently from Grishchuk's \cite{grishchuk},
$\zeta_E \equiv \frac{a(\tau_H)}{a(\tau_E)}$,
$\zeta_2 \equiv \frac{a(\tau_E)}{a(\tau_2)}$,
$\zeta_s \equiv \frac{a(\tau_2)}{a(\tau_s)}$,
and $ \zeta_1 \equiv \frac{a(\tau_s)}{a(\tau_1)}$.
With these specifications,
 the functions
$a(\tau)$ and $a'(\tau)/a(\tau)$ are fully determined.
In particular,
$a'(\tau)/a(\tau)$ rises up during the accelerating stage ,
instead of decreasing as in the matter-dominated stage.
This  causes  the modifications to the spectrum
of  relic gravitational waves.

\begin{center}
{\em\Large 4. The Spectrum of Gravitational Waves.}
\end{center}

The power spectrum $h( k, \tau)$ of  relic  gravitational waves is defined
by the following equation
\be
\int^{\infty}_0 h^2(k,\tau)\frac{dk}{k}  \equiv <0|h^{ij}
( {\bf x},\tau) h_{ij}( {\bf x},\tau) |0>,
\ee
where the right-hand-side is the vacuum expectation value of the  operator $ h^{ij} h_{ij} $.
Substituting Eq.(4) into the above,
and taking the contribution from each polarization to be  the same,
one reads the power spectrum
\be
h(k,\tau) = \frac{4l_{Pl}}{\sqrt{\pi}}k|h_k(\tau)|
\ee
Once the mode function  $h_k(\tau)$ is given, the spectrum $h(k,\tau)$ follows.

The initial condition is taken to be during the inflationary stage.
For a given wave number $k$,
its wave  crossed over the horizon at a time  $\tau_i$,
i.e.
when the wave length $\lambda_i = 2\pi a(\tau_i)/k$ is equal to $1/H(\tau_i)$,
the Hubble radius at time $\tau_i$.
Eq.(15) yields
$H(\tau_i) = l_0 |\tau_i|^{2+\beta}/|1+\beta|$,
and for the exact de Sitter expansion with $\beta =-2$ one has $H(\tau_i)=l_0$.
Note that a different $k$ corresponds to a different time $\tau_i$.
Now choose the initial condition of the mode function $h_k(\tau)$ as
\be
|h_k(\tau_i)| = \frac{1}{a(\tau_i)}.
\ee
Then the initial  amplitude of the power spectrum is
\be
h(k, \tau_i) = 8\sqrt{\pi} \frac{l_{Pl}}{\lambda_i}.
\ee
From  $ \lambda_i = 1/H(\tau_i) $ it follows that
$ \frac{a'(\tau_i)}{a(\tau_i)} = \frac{k}{2\pi} $.
So the initial  amplitude of the power spectrum is
\be
h(k, \tau_i) = A(\frac{k}{k_H})^{2+\beta} ,
\ee
where the constant
\be
A=8\sqrt{\pi}b
\frac{l_{Pl}}{l_0}.
\ee
For the case of  $\beta=-2$
the initial spectrum is independent of $k$.
The power spectrum for the primordial perturbations of  energy density
is $P(k)\propto  |h( k, \tau_H)|^2 $, and its spectral index $n$ is defined as
$P(k) \propto  k^{n-1}$.
Thus one reads off the relation $n = 2\beta +5$.
The exact de Sitter  expansion with $\beta = -2$ will yield
the so-called scale-invariant spectral index $n=1$.

Once the initial spectrum is specified,
we can derive the spectrum $h(k, \tau_H)$ at the present time $\tau_H$.

For $k  \leq k_E$,
the wave lengths $\frac{2\pi a(\tau_H)}{k}$ in this range are  even greater
than  the present  Hubble radius $l_H$,
one has $k< a'/a$
throughout  the whole expansion up to the present,
 so the  amplitude remains the same constant
as the initial one in Eq.(28):
\be
h( k, \tau_H) = A(\frac{k}{k_H})^{2+\beta}, \, \,\,\, k\leq  k_E.
\ee

During the whole period inside the barrier,
the  spectral  amplitude remains  $h(k,\tau_i)  $ approximately until
the wave  leaves the barrier and begins to decrease as $1/a(\tau)$.
Let $a_{**}(k)$ be the scale factor at this moment.
For those very long wavelength modes with $k_E \leq  k    \leq  k_H$,
during the  current epoch of accelerating expansion,
$h_k(\tau)$  stops decreasing as soon as the barrier $a'/a$
is higher than  $k$  at a time $\tau(k)$ earlier than $\tau_H$,
so $h_k(\tau)$  has decreased by a factor $\frac{a_{**}(k)}{a(\tau(k))}$,
and the amplitude of the present spectrum  is  given by
\be
h( k, \tau_H)= A(\frac{k}{k_H})^{2+\beta}  \frac{a_{**}(k)}{a(\tau(k))}.
\ee
The decreasing factor is written as
\[
 \frac{a_{**}(k)}{a(\tau(k))} =  \frac{a_{**}(k)}{a(\tau_E)} \frac{a(\tau_E)}{a(\tau(k))}.
\]
During the matter-dominated stage one has
$ a_{**}(k)\propto \tau^2 \propto k^{-2} $ and
$ a(\tau_E)\propto \tau_E^2 \propto k_E^{-2}  $,
and during the accelerating stage one has
$  a(\tau_E)\propto \tau_E^{-1} \propto k_E  $ and
$ a(\tau(k)) \propto \tau^{-1} \propto k $,
so the decreasing factor is
\[
 \frac{a_{**}(k)}{a(\tau(k))} =  (\frac{k_E}{k})^3.
\]
Thus one gets
\be
h( k,\tau_H) = A(\frac{k}{k_H})^{2+\beta} (\frac{k_E}{k})^3  A(\frac{k}{k_H})^{\beta -1}\frac{1}{(1+z_E)^3}, \,\,\,\,\, k_E \leq   k \leq k_H.
\ee
where the relation
$ \frac{k_E}{k_H}   = \frac{1}{ 1+z_E }$ has been used.
The spectrum has a rather stiff slope with the power-law index $\beta-1$.
The occurrence of this segment of power spectrum
is a new feature of the model of accelerating expansion
that is absent in the decelerating model.
The wave lengths corresponding to this $(k_E,\, k_H)$ segment  are very long,
 comparable to the present Hubble radius,
and can only possibly  be observed through the CMB anisotropies
at low multipoles.

For  all the wave number  $k>k_H$,
as soon as the waves leave the barrier at $a_{**}(k)$,
the modes $h_k(\tau)$ have been decreasing all the way
up to the present time $\tau_H$.
So it has been reduced by a factor $\frac{a_{**}(k)}{a(\tau_H)}$,
and   the amplitude of present spectrum is given by
\be
h(k,\tau_H)  = A(\frac{k}{k_H})^{2+\beta} \frac{a_{**}(k)}{a(\tau_H)}.
\ee
We use this formula to obtain
 the following result for the spectrum of the gravitational waves
in the remaining range of wave numbers.

For $k_H  \leq  k  \leq  k_2$,  the wave number does not hit the barrier,
so
\[
\frac{a_{**}(k)}{a(\tau_H)} \frac{a_{**}(k)}{a(\tau_E)} \frac{a(\tau_E)}{a(\tau_H)}
= (\frac{k_E}{k})^2  \frac{1}{(1+z_E)},
\]
one obtains
\be
h(k,\tau_H)
=A(\frac{k}{k_H})^{\beta} \frac{1}{(1+z_E)^3}, \,\,\,\, k_H  \leq   k  \leq  k_2 .
\ee
The spectrum in this interval  differs from
that of the matter-dominated model by an extra factor
$ \frac{1}{(1+z_E)^3}=  \frac{\Omega_m}{\Omega_{\Lambda}} \sim 0.43$.
The wave-lengths in this range  are very long,
but are still shorter than $l_H$.
The spectrum in this interval may contribute to,
and, therefore, have its imprints in  CMB anisotropies.
Let us estimate the value $k_2$.
Assuming that the equality of  radiation and  matter occurred
at the redshift $z_2 \simeq  3454$, as indicated by the  WMAP observation \cite{spergel},
one has
\[
\frac{k_2}{k_E } =  (\frac{ a(\tau_E) }{  a(\tau_2)  })^{\frac{1}{2}}
   = (\frac{ a(\tau_H) }{  a(\tau_2)  })^{\frac{1}{2}}
   (\frac{ a(\tau_E) }{  a(\tau_H)  })^{\frac{1}{2}}
   \simeq  \sqrt{3454}\frac{1}{\sqrt{1.33}}
\simeq  51 .
\]

For $k_2\leq   k\leq  k_s$, the calculation is similar to the previous case with  the result
\ba
h(k,\tau_H)  &  =  &  A(\frac{k}{k_H})^{ 1+ \beta} (\frac{k_H}{k_2} ) \frac{1}{(1+z_E)^3}
         , \,\,\,\,\,  k_2\leq  k\leq  k_s.
\ea
Again the extra factor $1/(1+z_E)^3$ appears.
Notice that this range of frequency covers  the one
that the detectors of LIGO and LISA are sensitive.

For $k_s\leq  k \leq k_1$, the calculation yields
\ba
h(k,\tau_H)
   & =   &   A(\frac{k}{k_H})^{1+\beta-\beta_s}
(\frac{k_s}{k_H}) ^{\beta_s}
\frac{k_H}{k_2}
\frac{1}{(1+z_E)^3}, \,\,\,\,\,   k_s\leq   k \leq  k_1.
\ea
It also contains an extra factor $\frac{1}{(1+z_E)^3} $.
This is the high frequency range which is still beyond the current detection.

\begin{center}
{\em\Large  5. Determining the Parameters.}
\end{center}

To completely determine the spectrum, we also need to specify  the values of  $\nu_1$,
$A$, $\nu_s$, $\beta_s$, that appear in the expressions for the spectrum $h(k, \tau_H)$.
Since the wave number is proportional to the frequency,
$k \propto \nu $, the ratios of the wave numbers can be replace by
that of the frequencies, e.g., $k/k_H =\nu/\nu_H$, etc., in the
above formulae of the spectrum $h(k, \tau_H)$. The Hubble
frequency $\nu_H= 1/l_H = H \simeq 2\times 10^{-18}$Hz.

An estimate of the highest frequency $\nu_1$ can be made
 as given in \cite{ grishchuk}.
From the expression (36)
one has
\be
 h(k_1 ,\tau_H)  = 8\sqrt{\pi} \frac{l_{Pl}}{l_H}
(\frac{\nu_1}{\nu_H}) = 8\sqrt{\pi} \frac{l_{Pl}}{l_1}  ,
\ee
where the expressions  (23) for $b/l_0$  and (29) for  $A$ have been used.
The spectral  energy density parameter $\Omega_g(\nu) $ of the
gravitational waves is defined through the relation
$\rho_g/\rho_c =\int \Omega_g(\nu) \frac{d\nu}{\nu}$,
where $\rho_g$ is the
energy density of the gravitational waves and $\rho_c$ is the
critical energy density. One reads
\[
 \Omega_g(\nu) = \frac{\pi^2}{3}h^2(k, \tau_H)(\frac{\nu}{\nu_H})^2   .
 \]
If it is imposed that at the highest frequency $\nu_1$ the value
$\Omega_g(\nu_1) $ not exceed the level of $10^{-6}$, as required
by the rate of the primordial nucleogenesis, then one gets
$  \nu_1 =3\times 10^{10}Hz  $.

Next let us  estimate the overall factor $A$ in the spectrum $h(k,\tau_H)$.
If  the CMB anisotropies at low multipoles
are induced by the gravitational waves,
or,  if the contributions from the gravitational waves and
from the density perturbations are of the same order of magnitude,
 we may assume $\Delta T/T \simeq h(k, \tau_H)$.
This will  determine  $A$.
The observed CMB anisotropies \cite{spergel} at lower multipoles
is $\Delta T/T \simeq 0.37  \times 10^{-5}$ at $l \sim 2$,
which corresponds to the largest scale anisotropies that have observed so far.
Taking this to be the perturbations at the Hubble radius $1/H$ yields
\be
h(k_H, \tau_H) = A \frac{1}{ (1+z_E)^3 }    = 0.37  \times 10^{-5}.
\ee
Actually the observed $\Delta T/T $  varies with $l$,
for instance,  $\Delta T/T \simeq 1.04  \times 10^{-5}$ at $l \sim 10$,
which would give an  outcome for $A$  greater than that
in Eq.(38) by a factor $\sim 2.8$.

However, there is a subtlety here on the interpretation of $\Delta T/T $  at low multipoles,
whose  corresponding scale is very large $\sim  l_H$.
At present the Hubble radius is $l_H  $, and the Hubble diameter is $2l_H$.
On the other hand,
the smallest characteristic wave number is  $k_E$,
whose corresponding physical wave length  at present is
$\frac{2\pi a(\tau_H)}{k_E}  = l_H(1+z_E)   \simeq  1.32 \, l_H$,
which is within the Hubble diameter $2l_H$,
and is theoretically observable.
So,  instead of (38),
if  $\Delta T/T \simeq 0.37  \times 10^{-5}$ at $l \sim 2$
were taken as   the amplitude of spectrum at  $ \nu_E$,
one would have
$
h(k_E, \tau_H) = A \frac{1}{ (1+z_E)^{2+\beta} } = 0.37  \times  10^{-5},
$
yielding  a smaller $A$  than that in Eq.(38)
by a factor  $ (1+z_E)^{1-\beta} \sim 2.2$.

We now check the range of $\beta$ allowed.
During the inflationary expansion when the $k$-mode  wave
entered the barrier with $\lambda_i= 1/H(\tau_i)$, from which it follows that
$
\lambda_i = \frac{l_0}{b} (\frac{k_H}{k})^{2+\beta}.
$
For the classical treatment of the background gravitational field
to be valid, this wave length  should be greater than the Planck
length, $\lambda_i > l_{Pl}  $, so
\[
( \frac{\nu}{\nu_H}  )^{2+\beta}  < \frac{8\sqrt{\pi}}{A} .
\]
At the highest frequency $\nu =\nu_1$, this yields
a constraint
\be
\beta <  -2 + \ln{(\frac{8\sqrt{\pi}}{A} )} /\ln{(\frac{\nu_1}{\nu_H}   ) },
\ee
which depends on $A$.
Thus, for $A$ given in (38), one obtains  the upper limit $ \beta <-1.78$.
We remark that this range  is larger than that
in the decelerating  model \cite{grishchuk},
which allows for only $ \beta <-1.9$.

Finally we give an estimate of the allowed values of $\beta_s$ and $\nu_s$.
Plugging $b/l_0$ given by (23) into $A=8\sqrt{\pi}b\frac{l_{Pl}}{l_0}$,
using $ \nu_2  / \nu_H = 58.8 $ and   $l_H/l_{Pl} =1.238\times 10^{61}$,  one has

\be
1.484\times 10^{58} \frac{A}{(1+z_E)^3}
= (\frac{\nu_1}{\nu_H})^{-\beta}
   (\frac{\nu_1}{\nu_s})^{\beta_s}  .
\ee
Given a set values of $A$ and $\beta$,
one can take $\nu_s$ and $\beta_s$ to satisfy this relation.
From Eq.(36) it is seen that, for a fixed $\nu_s$,
a smaller $\beta_s$ tends to increase slightly  the amplitude  $h(\nu, \tau_H)$.
For definiteness, we take $\nu_s = 10^8$ Hz as an example in the following.
For  $A$ given in (38),
taking $\beta =-1.8$,  $-1.9$,
then $\beta_s =  0.598$,   $ -0.552$, respectively.

\begin{figure}
\centerline{\includegraphics[width=10cm]{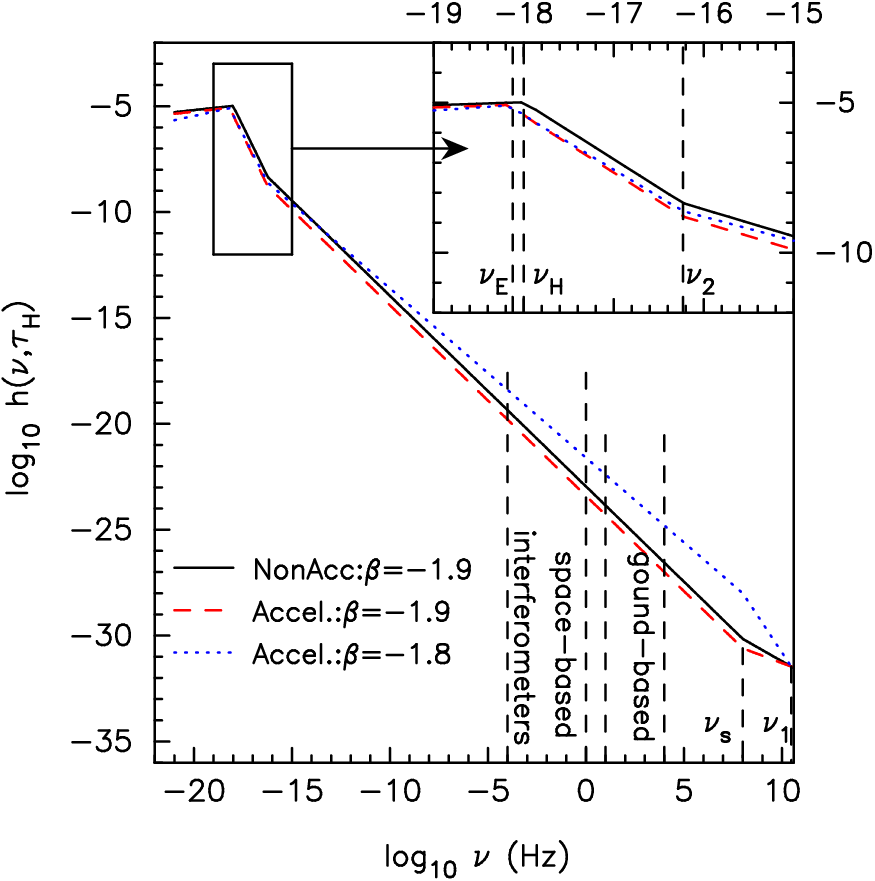}} \caption{The
spectrum $h(k,\tau_H)$ as a function of the frequency $\nu$
 in the present Universe in both accelerating and decelerating expansion.}
\end{figure}

\begin{figure}
\centerline{\includegraphics[width=9cm]{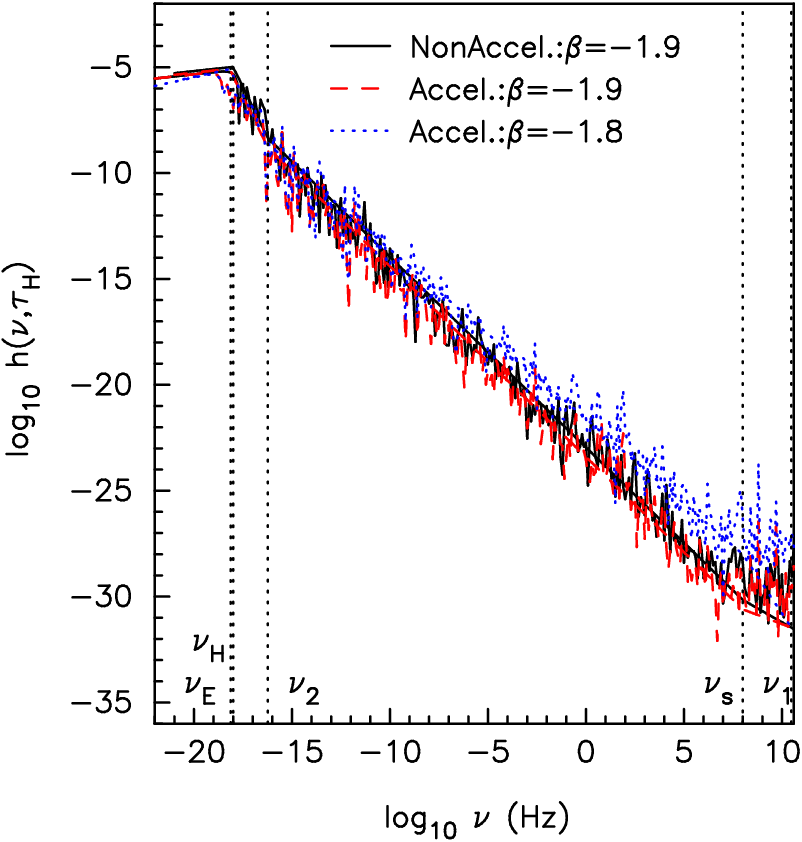}} \caption{The
spectrum $h(k,\tau_H)$ from the numerical calculation.}
\end{figure}

\begin{figure}
\centerline{\includegraphics[width=9cm]{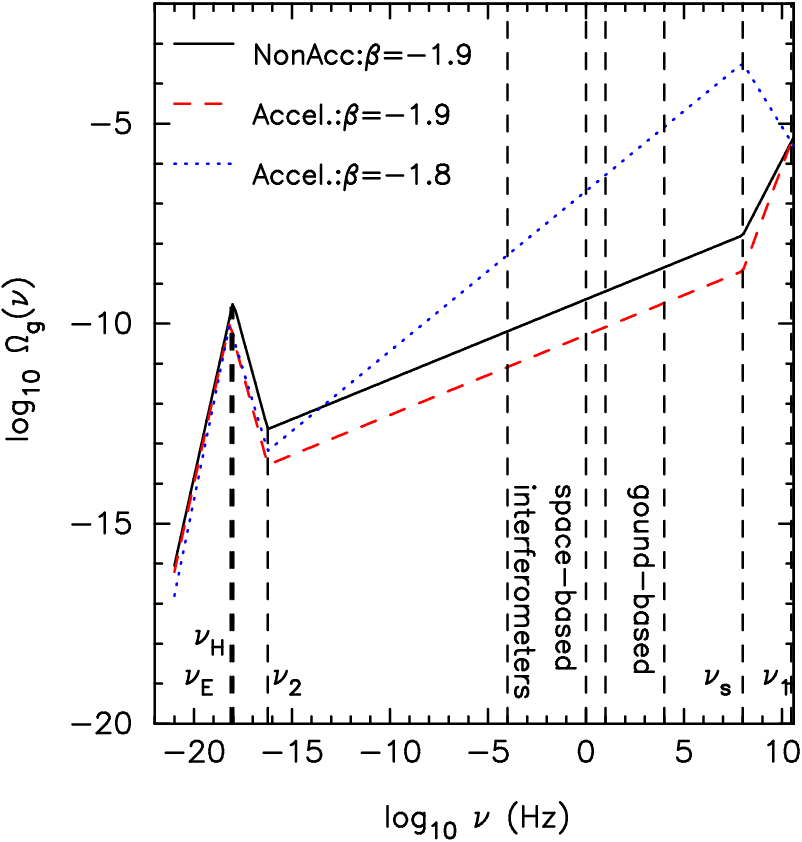}} \caption{The
energy spectrum $\Omega_g(\nu)$
 in the present Universe in both accelerating and decelerating expansion.}
\end{figure}

With these results we plot
the spectrum $h(k, \tau_H)$ vs the frequency $\nu$
for two values of parameter $\beta = -1.9,\, -1.8$ in Figure 1.
Besides, for convenience of comparison,
the spectrum for the decelerating  model
is also plotted for $\beta = -1.9$.
It is clearly seen that, for the same parameter $\beta= -1.9$
in the whole range $\nu \ge \nu_H$,
the accelerating model has an amplitude
lower by a factor of $\sim 2.2$ than that of the decelerating model.
This is due to  the extra factor $\frac{1}{(1+z_E)^3} = \frac{\Omega_m}{\Omega_{\Lambda}}$,
as has been demonstrated in the previous expressions (34), (35), and (36).
Interestingly, the larger value $\beta= -1.8$
gives a flatter spectrum $h(\nu,\tau_H)$ with an overall higher  amplitude
in the range $\nu \ge  \nu_2$.
In particular, in the higher  frequency range
$(10^{-4}\sim \, 0)$ Hz  covered by LISA,
the amplitude is $h\simeq 10^{-19} \sim 10^{-22}$ for the $\beta= -1.8$ case,
about $10$ to $10^2$ higher than the $\beta= -1.9$ case.
In the even higher frequency range  $(10\sim \,10^{4})$ Hz covered by  LIGO,
the amplitude is $h\simeq 10^{-23} \sim 10^{-25}$ for the $\beta= -1.8$ case,
also $10$ to $10^2$ higher than the $\beta= -1.9$ case.
Thus an inflationary model of larger index $n=2\beta+5$
will predict  a stronger signal of relic gravitational waves
in higher frequencies.

The recent second  science run of the LIGO interferometric detectors \cite{r-abbott}
gives a sensitivity $3\times 10^{-24}$ to $3\times 10^{-23}$ in the frequency range
$(10^2,\, 10^3)$Hz.
The  best sensitivity is given by the 4km arm L1 detector located at Livingstone,
which is about $3\times 10^{-24}$ near a frequency $\sim 300$Hz.
Our  calculation for the $\beta=-1.9$ case yields  an amplitude,
$h \simeq10^{-26}$,
 much smaller
than this sensitivity of the second run.
However, the most interesting case is the model of $\beta =-1.8$,
in which the amplitude  of the gravitational waves
is just to fall into the sensitivity of the L1 detector.
Since the second  run of LIGO has not observed any signal
of stochastic gravitational waves in this frequency range,
we arrive at a constraint
$
\beta \leq -1.8
$
 on the model parameter of the inflationary expansion.
 Note that this constraint is also consistent with
  $\beta < -1.78$,  as has been imposed,through Eq.(38),  from
  the observed CMB anisotropies of WMAP.

As a double check,
we have also numerically solved the differential equation (6) of the gravitational waves,
and have found the resulting power spectrum  $h(\nu, t)$  from the numerical solution,
which is plotted in Figure 2.
Since the numerical result that we have plotted
carries the oscillating factor $\cos (2\pi\nu (t-t_{\nu}))$,
so its curve shows an extra small zigzag, as expected.
Moreover, for the range of small frequencies,  $\nu <\nu_H$,
$\cos (2\pi\nu (t-t_{\nu}))\simeq 1$, the oscillating amplitude is very small,
confirming an observation made by ref. \cite{grishchuk}.
Except for  this oscillating effect, the numerical result of the spectrum in Figure 2
agree with the analytical one in Figure 1.

We also plotted the  spectral energy density $\Omega_g(\nu)$ from the analytic solution in Figure 3.
It is similar to the known result, except for the obvious distortions
caused by the acceleration of the Universe expansion in the low frequency range .

\begin{center}
{\em\Large  6. Conclusion.}
\end{center}

We have presented a calculation of the spectrum of  relic
gravitational waves in the present Universe in accelerating expansion.
The recent WMAP result  of $\Delta T/T $ has been used to normalize
the amplitude of relic gravitational waves.
In comparison with the decelerating  models,
the spectrum has been modified
due to the particular form of the barrier function $a'/a$
during the acceleration of current expansion.
Specifically, in very low frequency range $( < \nu_H  ) $
the spectrum has been  changed  with
the peak of spectrum being  now located at $\nu_E$,
and there appears  a new segment of spectrum from $\nu_E $ to $\nu_H$.
These very long wave length features can be only be possibly detected
by the CMB anisotropies at low multipoles.
In the higher frequency range $ ( \geq \nu_H   )$
the spectral amplitude acquires an overall  factor $\frac{\Omega_m}{\Omega_{\Lambda}}$
as compared with the decelerating model.
This higher frequency range is pertinent to the detection projects such as LISA and LIGO.
In addition, using the  $\Delta T/T $ at low multipoles for  normalization
gives  a larger upper limit  $\beta < -1.78$.
A larger value of $\beta$ yields  a flatter spectrum
$h(\nu,\tau_H)$ as a function of $\nu$,
producing  more power on the higher frequencies.
The resulting sensitivity of the second scientific run of the  LIGO detectors,
has put a restriction on the model parameter $\beta \leq -1.8$.

ACKNOWLEDGMENT: Y. Zhang's research work has
been supported by the Chinese NSF (10173008) and by NKBRSF (G19990754).
Y.F. Yuan is supported by the Special Funds for Major State Research Projects.


\end{document}